\documentclass[12pt,a4paper]{article}
\usepackage[utf8]{inputenc}
\usepackage{amsmath}
\usepackage{amsfonts}
\usepackage{amssymb}

\usepackage{tabularx}

\usepackage{graphicx}
\usepackage{caption}
\usepackage{float}
\usepackage[left=1in,right=1in,top=1in,bottom=1in]{geometry}
\usepackage{pdfcomment}
\usepackage{cite}
\def\th{\mathrm{th}}
\def\ahat{\hat{a}}
\def\Aphat{\hat{\textbf{A}}_p}
\def\Ahat{\hat{\mathcal{A}}}
\def\Ap{\textbf{A}_p}

\def\pr{\mathrm{pr}}
\def\C{\mathbf{C}}
\def\Thetav{\boldsymbol{\Theta}}
\def\noisev{\boldsymbol{N}}
\def\Omegav{\boldsymbol{\Omega}}
\def\etal{et al.~}
\DeclareMathOperator*{\argmax}{arg\,max} 

\title{A no-gold-standard technique to objectively evaluate quantitative imaging methods using patient data: Theory}
\author{{Jinxin Liu$^{1*\S}$, Ziping Liu$^{2*}$, Joyce Mhlanga$^{3}$, Barry A. Siegel$^{3}$, Abhinav K. Jha$^{2,3,\dagger}$} \\
	 $^{1}$Department of Electronic Engineering, Tsinghua University, Beijing, China \\
	 $^{2}$Department of Biomedical Engineering, Washington University in St. Louis, USA \\ 
	 $^{3}$Mallinckrodt Institute of Radiology, Washington University in St. Louis, USA \\
	 $^{*}$Contributed equally \\
	 $^\S$ Work conducted during summer internship at Washington University \\
	$^{\dagger}$Email: a.jha@wustl.edu
}

\begin{document}
\date{}
\maketitle

\begin{abstract}
Objective evaluation of quantitative imaging (QI) methods using measurements directly obtained from patient images is highly desirable but hindered by the non-availability of gold standards. To address this issue, statistical techniques have been proposed to objectively evaluate QI methods without a gold standard. These techniques assume that the measured and true values are linearly related by a slope, bias, and normally distributed noise term, where it is assumed that the noise term between the different methods is independent. However, the noise could be correlated since it arises in the process of measuring the same true value. Further, the existing methods assume a linear relationship between the true and measured values. To address this issues, we propose theory for a new no-gold-standard evaluation (NGSE) technique. This technique models a general polynomial relationship between true and measured values and models the noise as a multivariate normally distributed term, characterized by a covariance matrix. We derive a maximum-likelihood-based technique that, without any knowledge of the true QI values, estimates these polynomial terms and the elements of the covariance matrix. These are then used to rank the methods on the basis of precision of the measured QI values. This derivation demonstrates the mathematical premise behind the proposed NGSE technique. 
\end{abstract}

The focus of this document is to provide the theoretical formalism for a no-gold-standard evaluation (NGSE) technique. 
The formalism builds upon theory originally proposed as the regression-without-truth (RWT) technique  \cite{Kupinski:02, Hoppin:02, Hoppin:03, Kupinski:06}.
A quantitative imaging method measures a certain true quantitative value. 
The goal of the RWT technique was to evaluate different quantitative imaging methods on the task of measuring this true value even in the absence of the true value. 
The basic idea is that even though the true values are not known, the measured values are the result of a specific image-formation and quantification process that is applied to the true values. Thus, the measured and true values must be mathematically related. The RWT technique assumed that this relationship was linear, characterized by a slope, bias, and noise parameters. They demonstrated that, even in the absence of any ground truth, these parameters could be estimated using a maximum-likelihood technique. 

The RWT technique was applied Jha \etal and Buvat \etal to evaluate QI techniques for diffusion MR images \cite{Jha:12:pmb} and cardiac cine MR images \cite{Lebenberg:12}, respectively. The technique was then extended to a larger range of QI tasks \cite{Jha:16:PMB}. A mathematical intuition for this technique is provided in Jha \etal \cite{Jha:17:JMI}.
The results in all these studies demonstrated that NGSE techniques provide reliable evaluation of QI methods provided the assumptions made by the technique were satisfied. However, an important assumption made by existing NGSE techniques is that the noise component of the relationship between the different methods is independent for the different QI methods. Note that the noise with the different methods arises in the process of measuring the same true value, and thus could be correlated. 
Another assumption is that of linearity between the true and measured values, which may again be violated. 
We propose theoretical formalism for a new NGSE technique that does not make these assumptions. 

We derive the maximum-likelihood (ML) solution for the parameters that describe the relationship between the true and measured values, given the measured values from $P$ patient images using $K$ different quantitative imaging (QI) methods. 
In particular, we show that computing this ML solution does not require any knowledge of the true quantitative values. The formalism is presented for the more general case of a polynomial relationship between the true and measured QI values, although in the validation studies in this manuscript, we consider only the case where the relationship between the true and measured values is linear. 

Denote the true quantitative value for the $p^{\th}$ patient by $a_p$ and the measured quantitative value using the $k^{\th}$ QI method by $\ahat_{p,k}$. 
Let the relationship between the true and measured values be of the $M^{\th}$ order. 
The relationship can then be given by:
\begin{equation}
\ahat_{p,k} = u_{k,M} a_p^M + \ldots + u_{k,1} a_p + u_{k,0} + \epsilon_{p,k} \label{eq:poly_rel_est_true}
\end{equation}
Since the process of imaging and quantifying the measurement is a sequence of several random processes, using the central limit theorem, we assume that the noise term is normally distributed. 
For the values measured using the $K$ QI methods from the $p^{\th}$ patient, we can write this relationship in vector form as 
\begin{equation}
\left[ 
\begin{array} {c}  
\ahat_{p,1} \\ 
\ahat_{p,2} \\ 
\vdots \\
\ahat_{p,K}  
\end{array}
\right] = 
\begin{bmatrix}
u_{1,M} & \ldots & u_{1,1} & u_{1,0} \\
u_{2,M} & \ldots & u_{2,1} & u_{2,0} \\
\vdots & \vdots & \vdots & \vdots \\
u_{K,M} & \ldots & u_{K,1} & u_{K,0} \\
\end{bmatrix}
\begin{bmatrix}
a_p^M \\
\vdots \\
a_p \\
1 
\end{bmatrix} 
+ 
\left[ 
\begin{array} {c}  
\epsilon_{p,1} \\ 
\epsilon_{p,2} \\ 
\vdots \\
\epsilon_{p,K}  
\end{array}
\right] 
\end{equation}
where $\Thetav$ denotes the matrix of coefficients. 
Denote the measured values for the $p^{\th}$  patient from all $K$ imaging methods be  $\Aphat$. Further, denote the matrix consisting of all the coefficients above by $\Thetav$, the vector consisting of the different order of the true values by $\Ap$ and the vector of noise terms as $\noisev_p$. Then the above equation can be written more compactly as:
	\begin{equation}
	\Aphat = \Thetav \Ap + \noisev_p
	\end{equation}

To account for the fact that the noise between the different methods could be correlated, we assume that the  
we assume that the random vector $\noisev_p$ is a zero-mean multivariate normally distributed noise term:
	\begin{equation}
	\noisev_p \sim \mathcal{N}(0, \C)
	\end{equation}
	Then let $\pr(x|y)$ denote the conditional probability of a random variable $x$ when $y$ is known. 
	\begin{equation}
	\begin{aligned}
	\pr(\Aphat | a_p, \Theta, \C) &= \mathcal{N}(\Theta \Ap, \C) \\ &= \dfrac{1}{\sqrt{(2\pi)^k \cdot \det\textbf{C}}} \exp\{-\dfrac{1}{2} (\Aphat - \Theta \Ap)^T\textbf{C}^{-1}(\Aphat - \Theta \Ap)\},
	\label{eq:pr_Ahat_ap}
	\end{aligned}
	\end{equation}
where $\mathcal{N}$ denotes the multivariate normal distribution with mean $\Thetav \Ap$ and covariance matrix $\C$.
This distribution depends on $a_p$, which is not known. 
To circumvent this issue, assume $a_p$ has been sampled from some distribution parameterized by a vector $\Omegav$. 
Then the joint distribution of  $\Aphat$  and $a_p$ can be written as
	\begin{equation}
	\pr(\Aphat, a_p | \Thetav, \C, \Omegav) = \pr(\Aphat | \Ap,\Thetav, \C) \pr( \Ap | \Omegav)
	\end{equation}
We can apply marginalization (i.e. averaging) on both sides over the random variable $a_p$, which yields
	\begin{equation}
	\pr(\Aphat| \Thetav, \C, \Omegav) = \int d a_p \pr(\Aphat | \Ap, \Thetav, \C) \pr( \Ap | \Omegav)
	\end{equation}
After the marginalization, the distribution of $\Aphat$ is no more dependent on $a_p$. Finally, assuming that the true values are independent of each other, the joint distribution of all the measurements from all the patients, denoted by $\Ahat = \{\Aphat, p = 1, 2, \ldots P \}$  can be written simply as the product of the individual distributions of $\Aphat$,  i.e. 
	\begin{equation}
	\begin{aligned}
	\pr(\Ahat| \Theta, \C, \Omegav) &= \prod_{p=1}^P \int d a_p \pr(\Aphat | \Ap,\Thetav, \C) \pr( \Ap | \Omegav),
	\label{eq:pr_Ahat}
	\end{aligned}
	\end{equation}
where $\pr(\Aphat | \Ap, \Thetav, \C)$ is given by Eq.~\eqref{eq:pr_Ahat_ap}.
Eq.~\eqref{eq:pr_Ahat} yields the likelihood of all the measurements, parameterized in terms of the linear-relationship parameters and the true distribution parameters, and with no dependency on the true value. We can thus estimate the parameters that maximize this likelihood, yielding the ML solution:
	\begin{equation}
	\{ \hat{\Thetav}, \hat{\C}, \hat{\Omega} \}_{\mathrm{ML}} = \argmax_{\Thetav, \Omegav, \C} \pr( \Ahat | \Thetav, \C, \Omegav),
	\end{equation}
where $\argmax_x f(x)$ is the value of $x$ at which the function $f(x)$  is maximized. 

Instead of maximizing the likelihood, we maximize the logarithm of the likelihood. This yields:
	\begin{equation}
	\begin{aligned}
	\{ \hat{\Thetav}, \hat{\C}, \hat{\Omegav} \}_{\mathrm{ML}} &= \argmax_{\Thetav, \Omegav, \C} \{-\ln(\pr( \Ahat | \Thetav, \C, \Omegav))\}
	\\&= \argmax_{\Thetav, \Omegav, \C} -\Sigma_{p=1}^P \ln(\int d a_p \pr(\Aphat | \Ap,\Thetav, \C) \pr( \Ap | \Omegav))
	\label{eq:est_ngs_param}
	\end{aligned}
	\end{equation}
The ML estimator has several properties that make it an optimal technique to estimate these parameters. In particular, if an efficient estimator exists, the ML estimator is efficient, i.e. unbiased and attains the lowest bound on the variance of any unbiased estimator (Cramer Rao bound). Further, asymptotic variances and covariances of these estimates can be obtained directly from the inverse of the Fisher information matrix.

The NGSE technique requires that the unknown distribution of the true values is expressed in a parametric form. 
For this purpose, we chose the beta distribution. 
This form provides the ability to model a wide variety of shapes of the true distribution, including  including symmetric, non-symmetric, negatively-skewed, strictly increasing, strictly decreasing, concave, convex and uniform distributions.
Further, it is able to incorporate the constraint that the true values obtained in QI applications are typically positive.
The beta distribution function for the true value $a_p$ can be expressed in terms of the parameters $(\alpha, \beta )$ as follows:
	\begin{equation}
	\pr(a_p | \alpha, \beta) = \dfrac{(a_p)^{\alpha - 1}(1-a_p)^{\beta  -1}}{B(\alpha,\beta)},
	\label{eq:pr_beta_dist}
	\end{equation}
where $B(\alpha,\beta)$ denotes the beta function
	\begin{equation}
	B(\alpha,\beta) = \int_{0}^{1} u^{\alpha - 1}(1-u)^{\beta - 1}du
	\label{eq:beta_func}
	\end{equation}
In the NGSE formalism described above, $\Omega = (\alpha, \beta)$.

\bibliographystyle{unsrt}
\bibliography{refs}

\begin{thebibliography}{1}

\bibitem{Kupinski:02}
M~A Kupinski, J~W Hoppin, E~Clarkson, H~H Barrett, and G~A Kastis.
\newblock {{E}stimation in medical imaging without a gold standard}.
\newblock {\em Acad. Radiol.}, 9:290--7, Mar 2002.

\bibitem{Hoppin:02}
J~W Hoppin, M~A Kupinski, G~A Kastis, E~Clarkson, and H~H Barrett.
\newblock {{O}bjective comparison of quantitative imaging modalities without
  the use of a gold standard}.
\newblock {\em IEEE Trans. Med. Imaging}, 21:441--9, May 2002.

\bibitem{Hoppin:03}
J.~W. Hoppin, M.~A. Kupinski, D.~W. Wilson, T.~E. Peterson, B.~Gershman,
  G.~Kastis, E.~Clarkson, L.~Furenlid, and H.~H. Barrett.
\newblock {Evaluating estimation techniques in medical imaging without a gold
  standard: experimental validation}.
\newblock {\em Proc. SPIE}, 5034:230--237, 2003.

\bibitem{Kupinski:06}
M~A Kupinski, J~W Hoppin, J~Krasnow, S~Dahlberg, J~A Leppo, M~A King,
  E~Clarkson, and H~H Barrett.
\newblock {{C}omparing cardiac ejection fraction estimation algorithms without
  a gold standard}.
\newblock {\em Acad. Radiol.}, 13:329--37, Mar 2006.

\bibitem{Jha:17:JMI}
Abhinav~K Jha, Esther Mena, Brian~S Caffo, Saeed Ashrafinia, Arman Rahmim,
  Eric~C Frey, and Rathan~M Subramaniam.
\newblock Practical no-gold-standard evaluation framework for quantitative
  imaging methods: application to lesion segmentation in positron emission
  tomography.
\newblock {\em J. Med. Imag.}, 4(1):011011, 2017.

\bibitem{Jha:12:pmb}
A.~K. Jha, M.~A. Kupinski, J.~J. Rodriguez, R.~M. Stephen, and A.~T. Stopeck.
\newblock {Task-based evaluation of segmentation algorithms for
  diffusion-weighted MRI without using a gold standard}.
\newblock {\em Phys. Med. Biol.}, 57(13):4425--4446, Jul 2012.

\bibitem{Lebenberg:12}
J.~Lebenberg, I.~Buvat, A.~Lalande, P.~Clarysse, C.~Casta, A.~Cochet,
  C.~Constantinides, J.~Cousty, A.~De~Cesare, S.~Jehan-Besson, M.~Lefort,
  L.~Najman, E.~Roullot, L.~Sarry, C.~Tilmant, M.~Garreau, and F.~Frouin.
\newblock {Nonsupervised ranking of different segmentation approaches:
  Application to the estimation of the left ventricular ejection fraction from
  cardiac cine MRI sequences}.
\newblock {\em IEEE Trans. Med. Imaging}, 31(8):1651--1660, Aug 2012.

\bibitem{Jha:16:PMB}
Abhinav~K Jha, Brian Caffo, and Eric~C Frey.
\newblock A no-gold-standard technique for objective assessment of quantitative
  nuclear-medicine imaging methods.
\newblock {\em Phys. Med. Biol.}, 61(7):2780, 2016.

\end{thebibliography}
\end{document}